\documentclass[prd,showpacs,showkeys,nofootinbib,twocolumn,eqsecnum]{revtex4-1}
\usepackage{amsmath,amsfonts,amssymb}
\usepackage[colorlinks]{hyperref}
\usepackage{bm}
\usepackage{epsf}
\usepackage{color}
\usepackage{dcolumn}
\usepackage{bm}
\everymath{\displaystyle}

\begin{document} 

\title{Generalized plane waves in Poincar\'e gauge theory of gravity}

\author{\firstname{Milutin}~\surname{Blagojevi\'c}}
\email{mb@ipb.ac.rs}
\affiliation{Institute of Physics, University of Belgrade,
Pregrevica 118, 11080 Belgrade, Serbia}

\author{\firstname{Branislav}~\surname{Cvetkovi\'c}}
\email{cbranislav@ipb.ac.rs}
\affiliation{Institute of Physics, University of Belgrade,
Pregrevica 118, 11080 Belgrade, Serbia}

\author{\firstname{Yuri N.}~\surname{Obukhov}}
\email{obukhov@ibrae.ac.ru}
\affiliation{Theoretical Physics Laboratory, Nuclear Safety Institute,
Russian Academy of Sciences, B. Tulskaya 52, 115191 Moscow, Russia}


\begin{abstract}
A family of exact vacuum solutions, representing generalized plane waves propagating on the (anti-)de Sitter background, is constructed in the framework of Poincar\'e gauge theory. The wave dynamics is defined by the general Lagrangian that includes all parity even and parity odd invariants up to the second order in the gauge field strength. The structure of the solution shows that the wave metric significantly depends on the spacetime torsion. 
\end{abstract}

 \maketitle

\section{Introduction}

The gauge principle, which was originally formulated by Weyl in the context of electrodynamics \cite{Weyl}, now belongs to the key concepts which underlie the modern understanding of dynamical structure of fundamental physical interactions. Development of Weyl's idea, most notably in the works of Yang, Mills and Utiyama \cite{YM,Utiyama}, resulted in the construction of the general gauge-theoretic framework for arbitrary non-Abelian groups of {\it internal} symmetries. Sciama and Kibble extended this formalism to the {\it spacetime} symmetries, and proposed a theory of gravity \cite{Sciama,Kibble} based on the Poincar\'e group --a semidirect product of the group of spacetime translations times the Lorentz group. The importance of the Poincar\'e group in particle physics strongly supports the Poincar\'e gauge theory (PGT) as the most appropriate framework for description of the gravitational phenomena.

The ``translational'' gauge field potentials (corresponding to the subgroup of the spacetime translations) can be consistently identified with the spacetime coframe field, whereas the ``rotational'' gauge field potentials (corresponding to the local Lorentz subgroup) can be interpreted as the spacetime connection. This introduces the Riemann--Cartan geometry on the spacetime manifold, since one naturally recovers the torsion and the curvature as the Poincar\'e gauge field strengths \cite{rmp,MAG,Blag,selected,reader,Mielke,HonnefH,Tra,Cartan,Goenner,essay} (``translational'' and ``rotational'' one, respectively). The gravitational dynamics in PGT is determined by a Lagrangian that is assumed to be the function of the field strengths, the curvature and the torsion, and the dynamical setup is completed by including a suitable matter Lagrangian. 

In the past, investigations of PGT were mostly focused on the class of models with quadratic parity symmetric Lagrangians of the Yang-Mills type, expecting that the results obtained for such a class should be sufficient to reveal essential dynamical features of the more complex general theory, for an overview see \cite{Obukhov:1989}. Recently, however, there has been a growing interest for the extended class of models with a general Lagrangian that includes both parity even and parity odd quadratic terms, see for instance \cite{Ho1,Ho2,Ho3,Diakonov,Baekler1,Baekler2}. An important difference between these two classes of PGT models is manifest in their particle spectra. Generically, the particle spectrum of the parity conserving PGT model consists of the massless graviton and eighteen massive torsion modes. The conditions for the absence of ghosts and tachyons impose serious restrictions on the propagation of these modes \cite{Hay2,Nev1,Nev2,Sez1,Sez2,Kuh}. In contrast, a recent analysis of the general PGT \cite{Karananas} shows that the propagation of torsion modes is much less restricted. This is a new and physically interesting dynamical effect of the parity odd sector.

Based on the experience stemming from general relativity (GR), it is well known that exact solutions play an important role in understanding gravitational dynamics. An important class of these solutions consists of the gravitational waves  \cite{curr,vdz,griff,exact,griff2}, one of the best known families of exact solutions in GR. For many years, investigation of gravitational waves has been an interesting subject also in the framework of PGT \cite{adam,chen,sippel,vadim,singh,babu,BC1,BC2,BC3,BC4}, as well as in the metric-affine gravity theory which is obtained in the gauge-theoretic approach when the Poincar\'e group is extended to the general affine symmetry group \cite{ppmag,dirk1,dirk2,king,vas1,vas2,vas3,pasic1,pasic2}. Noticing that dynamical effects of the parity odd sector of PGT are not sufficiently well known, recently one of us \cite{PGW} has studied exact plane wave solutions with torsion in vacuum, propagating on the flat background, for the case of the vanishing cosmological constant $\Lambda$. In another recent work \cite{BC5} complementary results have been obtained, when the generalized $pp$ waves with torsion were derived as exact vacuum solutions of the parity even PGT, but for the case of a nontrivial $\Lambda\ne 0$. In the present paper, we merge and extend these investigations by constructing the generalized plane waves with torsion as vacuum solutions of the general quadratic PGT with nonvanishing cosmological constant. The resulting structure offers a deeper insight into the dynamical role of the parity odd sector of PGT.

The paper is organized as follows. In the next Sec.~\ref{PG} we present a condensed introduction to the Poincar\'e gauge gravity theory, giving the basic definitions and describing the main structures; more details can be found in \cite{rmp,MAG,Blag,reader}. In Sec.~\ref{GW} we start with representing an (anti)-de Sitter spacetime as a gravitational wave and use the properties of the plane-fronted electromagnetic and gravitational waves discussed in \cite{ndim} to formulate an ansatz for the gravitational wave in the Poincar\'e gauge gravity. The properties of the resulting curvature and torsion 2-forms are studied. In Sec.~\ref{FE} the set of differential equations for the wave variables is derived. It is worthwhile to note that the functions which describe wave's profile satisfy a system of linear equations, even though the original field equations of the Poincar\'e gauge theory are highly nonlinear. Solutions of this system are constructed, and their properties are discussed. We demonstrate the consistency of the results obtained with the particle spectrum of the general Poincar\'e gauge gravity model. Finally, the conclusions are outlined in Sec.~\ref{DC}.

Our basic notation and conventions are consistent with \cite{MAG}. In particular, Greek indices $\alpha, \beta, \dots = 0, \dots, 3$, denote the anholonomic components (for example, of a coframe $\vartheta^\alpha$), while the Latin indices $i,j,\dots =0,\dots, 3$, label the holonomic components ($dx^i$, e.g.). The anholonomic vector frame basis $e_\alpha$ is dual to the coframe basis in the sense that $e_\alpha\rfloor\vartheta^\beta = \delta_\alpha^\beta$, where $\rfloor$ denotes the interior product. The volume 4-form is denoted $\eta$, and the $\eta$-basis in the space of exterior forms is constructed with the help of the interior products as $\eta_{\alpha_1 \dots\alpha_p}:= e_{\alpha_p}\rfloor\dots e_{\alpha_1}\rfloor\eta$, $p=1,\dots,4$. They are related to the $\vartheta$-basis via the Hodge dual operator $^*$, for example, $\eta_{\alpha\beta} = {}^*\!\left(\vartheta_\alpha\wedge\vartheta_\beta\right)$. The Minkowski metric $g_{\alpha\beta} = {\rm diag}(+1,-1,-1,-1)$. All the objects related to the parity-odd sector (coupling constants, irreducible pieces of the curvature, gravitational wave potentials, etc) are marked by an overline, to distinguish them from the corresponding parity-even objects.

\section{Basics of Poincar\'e gauge gravity}\label{PG}

The gravitational field is described by the coframe $\vartheta^\alpha = e^\alpha_i dx^a$ and connection $\Gamma_\alpha{}^\beta = \Gamma_{i\alpha}{}^\beta dx^i$ 1-forms. The translational and rotational field strengths read 
\begin{eqnarray}
T^\alpha &=& D\vartheta^\alpha = d\vartheta^\alpha +\Gamma_\beta{}^\alpha\wedge
\vartheta^\beta,\label{Tor}\\ \label{Cur}
R_\alpha{}^\beta &=& d\Gamma_\alpha{}^\beta + \Gamma_\gamma{}^\beta\wedge\Gamma_\alpha{}^\gamma.
\end{eqnarray}
As usual, the covariant differential is denoted $D$.

The gravitational Lagrangian 4-form is (in general) an arbitrary function of the geometrical variables:
\begin{equation}
V = V(\vartheta^{\alpha}, T^{\alpha}, R_{\alpha}{}^{\beta}).\label{lagrV}
\end{equation}
Its variation with respect to the gravitational (translational and Lorentz) potentials yields the field equations
\begin{eqnarray}
{\mathcal E}_\alpha &:=& {\frac{\delta V}{\delta\vartheta^{\alpha}}} = 
- DH_{\alpha}  +E_{\alpha} = 0, \label{dVt}\\ 
{\mathcal C}^\alpha{}_\beta &:=& {\frac{\delta V}{\delta\Gamma_\alpha{}^\beta}} 
= - DH^\alpha{}_\beta + E^\alpha{}_\beta\, = 0.\label{dVG}
\end{eqnarray}
Here, the Poincar\'e {\it gauge field momenta} 2-forms are introduced by
\begin{equation} 
H_{\alpha} := -\,{\frac{\partial V}{\partial T^{\alpha}}}\,,\qquad  
H^{\alpha}{}_{\beta} := -\,{\frac{\partial V}{\partial R_{\alpha}{}^{\beta}}}\,,\label{HH}
\end{equation}  
and the $3$--forms of the {\it canonical} energy--momentum and 
spin for the gravitational gauge fields are constructed as
\begin{eqnarray} 
E_{\alpha} := {\frac{\partial V}{\partial\vartheta^{\alpha}}} &=& e_{\alpha}\rfloor V 
+ (e_{\alpha}\rfloor T^{\beta})\wedge H_{\beta} \nonumber\\
&& + \,(e_{\alpha}\rfloor R_{\beta}{}^{\gamma})\wedge H^{\beta}{}_{\gamma},\label{Ea}\\
E^\alpha{}_\beta := {\frac{\partial V}{\partial\Gamma_\alpha{}^\beta}} &=&  
- \,\vartheta^{[\alpha}\wedge H_{\beta]}\,. \label{EE}
\end{eqnarray}

The field equations (\ref{dVt}) and (\ref{dVG}) are written here for the vacuum case. In the presence of matter, the right-hand sides of (\ref{dVt}) and (\ref{dVG}) contain the canonical energy-momentum and the canonical spin currents of the physical sources, respectively.

\subsection{Quadratic Poincar\'e gravity models}

The torsion 2-form can be decomposed into the 3 irreducible parts, whereas the curvature 2-form has 6 irreducible pieces. Their definition is presented in Appendix~\ref{irreducible}.

The general quadratic model is described by the Lagrangian 4-form that contains all possible quadratic invariants of the torsion and the curvature:
\begin{eqnarray}
V &=& {\frac {1}{2\kappa c}}\Big\{\Big(a_0\eta_{\alpha\beta} + \overline{a}_0
\vartheta_\alpha\wedge\vartheta_\beta\Big)\wedge R^{\alpha\beta} - 2\lambda_0\eta \nonumber\\
&& -\,T^\alpha\wedge\sum_{I=1}^3
\left[a_I\,{}^*({}^{(I)}T_\alpha) + \overline{a}_I\,{}^{(I)}T_\alpha\right]\Big\}\nonumber\\
&& - \,{\frac 1{2\rho}}R^{\alpha\beta}\wedge\sum_{I=1}^6 \left[b_I\,{}^*({}^{(I)}\!R_{\alpha\beta}) 
+ \overline{b}_I\,{}^{(I)}\!R_{\alpha\beta}\right].\label{LRT}
\end{eqnarray}
The Lagrangian has a clear structure: the first line is {\it linear} in the curvature, the second line collects {\it torsion quadratic} terms, whereas the third line contains the {\it curvature quadratic} invariants. Furthermore, each line is composed of the parity even pieces (first terms on each line), and the parity odd parts (last terms on each line). The dimensionless constant $\overline{a}_0 = {\frac 1\xi}$ is inverse to the so-called Barbero-Immirzi parameter $\xi$, and the linear part of the Lagrangian -- the first line in (\ref{LRT}) -- describes what is known in the literature as the Einstein-Cartan-Holst model. A special case $a_0 = 0$ and $\overline{a}_0 = 0$ describes the purely quadratic model without the Hilbert-Einstein linear term in the Lagrangian. In the Einstein-Cartan model, one puts $a_0 = 1$ and $\overline{a}_0 = 0$. 

Besides that, the general PGT model contains a set of the coupling constants which determine the structure of quadratic part of the Lagrangian: $\rho$, $a_1, a_2, a_3$ and $\overline{a}_1, \overline{a}_2, \overline{a}_3$, $b_1, \cdots, b_6$ and $\overline{b}_1, \cdots, \overline{b}_6$. The overbar denotes the constants responsible for the parity odd interaction. We have the dimension $[{\frac 1\rho}] = [\hbar]$, whereas $a_I$, $\overline{a}_I$, $b_I$ and $\overline{b}_I$ are dimensionless. Moreover, not all of these constants are independent: we take $\overline{a}_2 = \overline{a}_3$, $\overline{b}_2 = \overline{b}_4$ and $\overline{b}_3 = \overline{b}_6$ because some of terms in (\ref{LRT}) are the same in view of (\ref{T23})-(\ref{R36}).

For the Lagrangian (\ref{LRT}) from (\ref{HH})-(\ref{EE}) we derive the gauge gravitational field momenta 
\begin{eqnarray}
H_\alpha &=& {\frac 1{\kappa c}}\,h_\alpha\,, \label{HaRT}\\
H^\alpha{}_\beta &=& -\,{\frac {1}{2\kappa c}}\left(a_0\,\eta^\alpha{}_\beta + \overline{a}_0
\vartheta^\alpha\wedge\vartheta_\beta\right)  + {\frac 1\rho}\,h^\alpha{}_\beta,\label{HabRT}
\end{eqnarray}
and the canonical energy-momentum and spin currents of the gravitational field
\begin{eqnarray}
E_\alpha &=& {\frac {1}{2\kappa c}}\Big(a_0\,\eta_{\alpha\beta\gamma}\wedge R^{\beta\gamma} + 
2\overline{a}_0\,R_\alpha{}^\beta\wedge\vartheta_\beta \nonumber\\
&& -\,2\lambda_0\eta_\alpha + q^{(T)}_\alpha\Big) + {\frac 1\rho}\,q^{(R)}_\alpha,\label{EaRT}\\
E^\alpha{}_\beta &=& {\frac 12}\left(H^\alpha\wedge\vartheta_\beta 
- H_\beta\wedge\vartheta^\alpha\right).\label{EabRT}
\end{eqnarray}
For convenience, we introduced here the 2-forms which are linear functions 
of the torsion and the curvature, respectively, by
\begin{eqnarray}
h_\alpha &=& \sum_{I=1}^3\left[a_I\,{}^*({}^{(I)}T_\alpha) 
+ \overline{a}_I\,{}^{(I)}T_\alpha\right],\label{HlaQT}\\
h^\alpha{}_\beta &=& \sum_{I=1}^6\left[b_I\,{}^*({}^{(I)}\!R^\alpha{}_\beta) 
+ \overline{b}_I\,{}^{(I)}\!R^\alpha{}_\beta\right],\label{hR}
\end{eqnarray}
and the 3-forms which are quadratic in the torsion and in the curvature, respectively:
\begin{eqnarray}
q^{(T)}_\alpha &=& {\frac 12}\left[(e_\alpha\rfloor T^\beta)\wedge h_\beta - T^\beta\wedge 
e_\alpha\rfloor h_\beta\right],\label{qa}\\
q^{(R)}_\alpha &=& {\frac 12}\left[(e_\alpha\rfloor R_\beta{}^\gamma)\wedge h^\beta{}_\gamma 
- R_\beta{}^\gamma\wedge e_\alpha\rfloor h^\beta{}_\gamma\right].\label{qaR}
\end{eqnarray}
By construction, (\ref{HlaQT}) has the dimension of a length, $[h_\alpha] = [\ell]$, whereas the 2-form (\ref{hR}) is obviously dimensionless, $[h^\alpha{}_\beta] = 1$. Similarly, we find for (\ref{qa}) the dimension of length $[q^{(T)}_\alpha] = [\ell]$, and the dimension of the inverse length, $[q^{(R)}_\alpha] = [1/\ell]$ for (\ref{qaR}). 

The resulting {\it vacuum} Poincar\'e gravity field equations (\ref{dVt}) and (\ref{dVG}) then read:
\begin{eqnarray}
{\frac {a_0}2}\eta_{\alpha\beta\gamma}\wedge R^{\beta\gamma} + \overline{a}_0R_\alpha{}^\beta
\wedge\vartheta_\beta  && \nonumber\\ 
-\,\lambda_0\eta_\alpha + q^{(T)}_\alpha + \ell_\rho^2\,q^{(R)}_\alpha - Dh_\alpha &=& 0,\label{ERT1}\\
a_0\,\eta^\alpha{}_{\beta\gamma}\wedge T^{\gamma} + \overline{a}_0\left(T^\alpha\wedge\vartheta_\beta
- T_\beta\wedge\vartheta^\alpha \right) && \nonumber\\ 
+ \,h^\alpha\wedge\vartheta_\beta - h_\beta\wedge\vartheta^\alpha 
- 2\ell_\rho^2\,Dh^\alpha{}_\beta &=& 0.
\label{ERT2}
\end{eqnarray}
The contribution of the curvature square terms in the Lagrangian (\ref{LRT}) to the gravitational field dynamics in the equations (\ref{ERT1}) and (\ref{ERT2}) is characterized by the parameter
\begin{equation}
\ell_\rho^2 = {\frac {\kappa c}{\rho}}.\label{lr}
\end{equation}
Since $[{\frac 1\rho}] = [\hbar]$, this new coupling parameter has the dimension of the area, $[\ell_\rho^2] = [\ell^2]$.

\section{Gravitational waves in Poincar\'e gauge gravity}\label{GW}

Gravitational waves are of fundamental importance in physics, and recently the purely theoretical research in this area was finally supported by the first experimental evidence \cite{Abbott1,Abbott2,Abbott3}. A general overview of the history of this fascinating subject can be found in \cite{flan,schutz,CNN}.

\subsection{(Anti)-de Sitter spacetime as a wave}

Let us now discuss the four-dimensional manifold which can be viewed as an ``(anti)-de Sitter spacetime in the wave disguise''. As before \cite{PGW}, we use the same local coordinates which are divided into two groups: $x^i= (x^a, x^A)$, where $x^a = (x^0 = \sigma, x^1 = \rho)$ and $x^A = (x^2,x^3)$. Hereafter the indices from the beginning of the Latin alphabet label the coordinates $\sigma$ and $\rho$ parametrizing the wave rays, $a,b,c... = 0,1$, whereas the capital Latin indices, $A,B,C... = 2,3$, refer to coordinates $x^A$ on the wave front. 

The coframe 1-form is chosen as a direct generalization of the ansatz used in \cite{PGW,ndim}:
\begin{eqnarray}
\widehat{\vartheta}^{\widehat 0} &=& {\frac {q}{2p}}\left[(\widehat{U} + 1)d\sigma 
+ d\rho\right],\label{cofDS0}\\ 
\widehat{\vartheta}^{\widehat 1} &=& {\frac {q}{2p}}\left[(\widehat{U} - 1)d\sigma 
+ d\rho\right],\label{cofDS1}\\
\widehat{\vartheta}^{\widehat A} &=& {\frac 1p}\,dx^A,\qquad A = 2,3.\label{cofDS23}
\end{eqnarray}
Here the three functions are given by the following expressions:
\begin{eqnarray}
\widehat{U} &=& -\,{\frac {\lambda}{4}}\,\rho^2,\label{Uhat}\\
p &=& 1 + {\frac {\lambda}{4}}\,\delta_{AB}x^Ax^B,\label{p}\\ 
q &=& 1 - {\frac {\lambda}{4}}\,\delta_{AB}x^Ax^B.\label{q}
\end{eqnarray}
The constant parameter $\lambda$ is an arbitrary real number (which can be 
positive, negative or zero). As a result, the line element reads
\begin{equation}
ds^2 = {\frac 1{p^2}}\left\{q^2\left(d\sigma d\rho + \widehat{U}d\sigma^2\right) 
- \delta_{AB}dx^Adx^B\right\}.\label{dsDS}
\end{equation}

The key object for the description of the wave configurations is the wave 1-form. 
On the basis of the earlier results \cite{PGW}, we introduce a wave 1-form $k$ as
\begin{equation}
k := d\sigma = {\frac pq}\left(\widehat{\vartheta}^{\widehat 0} 
- \widehat{\vartheta}^{\widehat 1}\right).\label{kDS}
\end{equation}
By construction, we have $k\wedge{}^\ast\!k = 0$. As before, the wave covector is $k_\alpha = e_\alpha\rfloor k$. Its (anholonomic) components are thus $k_\alpha = {\frac pq}(1, -1, 0, 0)$ and $k^\alpha = {\frac pq}(1, 1, 0, 0)$. Hence, this is a null vector field, $k_\alpha k^\alpha = 0$.

The corresponding Riemannian connection $\widehat{\Gamma}_\beta{}^\alpha$ is determined from
\begin{equation}\label{notorDS}
d\widehat{\vartheta}^\alpha + \widehat{\Gamma}_\beta{}^\alpha\wedge\widehat{\vartheta}^\beta = 0,
\end{equation}
and it reads explicitly (recall that $a,b,\dots = 0,1$ and $A,B,\dots = 2,3$)
\begin{eqnarray}
\widehat{\Gamma}_{\widehat 0}{}^{\widehat 1} &=& \widehat{\Gamma}_{\widehat 1}{}^{\widehat 0}
= -\,{\frac {\lambda\rho}{2}}\,k,\label{rDS1}\\
\widehat{\Gamma}_B{}^a &=& {\frac pq}\,\widehat{\vartheta}^a\,e_B\rfloor d\Bigl({\frac qp}
\Bigr),\label{rDS2}\\
\widehat{\Gamma}_B{}^A &=& {\frac 1p}\left(\widehat{\vartheta}_B\,e^A\rfloor dp 
- \widehat{\vartheta}^A\,e_B\rfloor dp\right).\label{rDS3}
\end{eqnarray}
Substituting (\ref{Uhat})-(\ref{q}), we straightforwardly find the curvature:
\begin{equation}\label{curDS}
\widehat{R}_\beta{}^\alpha = \lambda\,\widehat{\vartheta}_\beta\wedge\widehat{\vartheta}^\alpha.
\end{equation}
Thus, the coframe and connection $(\widehat{\vartheta}^\alpha, \widehat{\Gamma}_\beta{}^\alpha)$, described by (\ref{cofDS0})-(\ref{cofDS23}) and (\ref{rDS1})-(\ref{rDS3}), represent the geometry of a torsionless (\ref{notorDS}) spacetime of constant curvature (\ref{curDS}). Depending on the sign of $\lambda$, we have either a de Sitter or an anti-de Sitter space.

We mark the corresponding geometrical quantities by the hat over the symbols. This geometry will be used as a starting point for the construction of the plane wave solutions in the Poincar\'e gauge gravity with nontrivial cosmological constant. 

It is worthwhile to note that the wave vector field $k$ is a null geodesic in this geometry:
\begin{equation}
k\wedge{}^\ast\!k = 0,\qquad k\wedge{}^\ast\!\widehat{D}k^\alpha = 0.\label{kkDS}
\end{equation}

\subsection{Generalized plane wave ansatz}

We will construct new gravitational wave solutions in Poincar\'e gauge gravity theory by making use of the ansatz for the coframe and for the local Lorentz connection
\begin{eqnarray}\label{cofDSPG}
\vartheta^\alpha &=& \widehat{\vartheta}^\alpha + {\frac U2}\,{\frac qp}k^\alpha\,k,\\
\Gamma_\alpha{}^\beta &=& \widehat{\Gamma}_\alpha{}^\beta 
+ {\frac qp}\left(k_\alpha W^\beta - k^\beta W_\alpha\right)k.\label{conDS}
\end{eqnarray}
Here the function $U = U(\sigma, x^A)$ determines the wave profile. The ansatz for the local Lorentz connection is postulated as a direct analogue of the construction used earlier in \cite{PGW}, and the vector variable $W^\alpha = W^\alpha(\sigma, x^A)$ satisfies the same orthogonality property, $k_\alpha W^\alpha = 0$, which is guaranteed by the choice
\begin{equation}\label{Wa0}
W^\alpha = \begin{cases}W^a = 0,\qquad\qquad\qquad a = 0,1, \\
W^A = W^A(\sigma, x^B),\qquad A = 2,3.\end{cases}
\end{equation}
Consequently, the generalized ansatz for the Poincar\'e gauge potentials -- coframe (\ref{cofDSPG}) and connection (\ref{conDS}) -- is described by the three variables $U = U(\sigma, x^B)$ and $W^A = W^A(\sigma, x^B)$. These should be determined from the gravitational field equations.

The ansatz (\ref{cofDSPG}) and (\ref{conDS}) can be viewed as a non-Riemannian extension of the Kerr-Schild-Kundt construction developed recently \cite{Gurses:2011,Gurses:2012,Gurses:2013,Gurses:2014} in general relativity and in modified gravity models. The original Kerr-Schild construction \cite{exact} in GR is underlain by the existence of preferred null directions. In our approach, the metric defined by the coframe (\ref{cofDSPG}) can be written in a typical Kerr-Schild form
\begin{equation}
g_{ij} = \widehat{g}{}_{ij} + \frac{q}{p}Uk_i k_j\, ,
\end{equation}
where $\widehat{g}_{ij}$ is the spacetime metric of the (anti)-de Sitter line element (\ref{dsDS}), and $k_i = \partial_i\rfloor k = \partial_i\rfloor d\sigma = (1, 0, 0, 0)$ is the null vector with respect to both $\widehat{g}{}_{ij}$ and $g_{ij}$. On the other hand, the orthogonality property of the vector $W^\alpha$ that defines the radiation piece of the connection (\ref{Wa0}), $k_\alpha W^\alpha=0$, ensures typical radiation structure of the Riemann-Cartan field strengths, the torsion and the curvature.

The line element for this ansatz has the same form (\ref{dsDS}), with a replacement
\begin{equation}
\widehat{U} \quad\longrightarrow\quad \widehat{U} + {\frac pq}\,U.\label{U2U}
\end{equation}
It is important to stress that the wave 1-form $k$ is still defined by (\ref{kDS}), which 
however can be recast into 
\begin{equation}
k = d\sigma = {\frac pq}\left(\vartheta^{\widehat 0} - \vartheta^{\widehat 1}\right).\label{kDSPG}
\end{equation}
Consequently, the anholonomic components of the wave covector $k_\alpha = e_\alpha\rfloor k$ still have the values $k_\alpha = {\frac pq}(1, -1, 0, 0)$ and $k^\alpha = {\frac pq}(1, 1, 0, 0)$. As before, this is a null vector field, $k_\alpha k^\alpha = 0$.

One may wonder why does the factor ${\frac qp}$ appear in the ansatz (\ref{cofDSPG}) and (\ref{conDS}). After all, it is always possible to absorb it by redefining $U$ and $W^A$. However, it is convenient to keep this factor explicitly by noticing that the combination ${\frac qp}k^\alpha = (1, 1, 0, 0)$ has the constant values. It becomes clear then that the following differential relations are valid:
\begin{equation}\label{dkDS1}
dk = 0,\qquad d\Bigl({\frac qp}k_\alpha\Bigr) = 0.
\end{equation}
Moreover, although $Dk_\alpha$ no longer vanishes, we find
\begin{equation}\label{dkDS2}
k\wedge D\Bigl({\frac qp}k_\alpha\Bigr) = k\wedge \widehat{D}\Bigl({\frac qp}k_\alpha\Bigr) = 0.
\end{equation}
Taking this into account, we straightforwardly compute the torsion 2-form
\begin{eqnarray}
T^\alpha = -\,k\wedge {\frac qp}k^\alpha\,\Theta,\label{torDS}
\end{eqnarray}
where we introduced the 1-form 
\begin{eqnarray}
\Theta = {\frac 12}\,\underline{d}\,U + W_\alpha\vartheta^\alpha,\label{THDS}
\end{eqnarray}
with the differential $\underline{d} := \vartheta^Ae_A\rfloor d = dx^A\partial_A$ that
acts in the transversal 2-space spanned by $x^A = (x^2, x^3)$. 

The structure of the torsion is qualitatively the same as in the case of the vanishing parameter $\lambda$, see \cite{PGW}. The structure of curvature is more nontrivial, though. A direct computation yields a 2-form
\begin{eqnarray}
R_\alpha{}^\beta =  \lambda\,\vartheta_\alpha\wedge\vartheta^\beta
-\,k\wedge{\frac qp}\left(k_\alpha \Omega^\beta - k^\beta \Omega_\alpha\right),\label{curDSPG}
\end{eqnarray}
where we introduced the vector-valued 1-form with the components
\begin{eqnarray}\label{OMDS}
\Omega^\alpha  = \begin{cases} \Omega^a = 0,\qquad\qquad\qquad\qquad a = 0,1, \\
\Omega^A = \widehat{D}W^A + {\frac {\lambda}{2}}\,U\vartheta^A,\qquad A = 2,3.\end{cases}
\end{eqnarray}
The transversal covariant derivative is defined by
\begin{equation}
\widehat{D}W^A = \underline{d}W^A + \widehat{\Gamma}_B{}^A\,W^B.\label{DW}
\end{equation}
Note that the Riemannian de Sitter connection (\ref{rDS3}) appears here (more exactly, the corresponding components of the Riemann-Cartan connection (\ref{conDS}) coincide with the Riemannian components: $\Gamma_B{}^A = \widehat{\Gamma}_B{}^A$). 

Let us describe the geometry of the transversal 2-space spanned by $x^A = (x^2, x^3)$ explicitly. The volume 2-form reads $\underline{\eta} = {\frac 12}\eta_{AB}\vartheta^A\wedge\vartheta^B = {\frac {1}{p^2}}dx^2\wedge dx^3$, where $\eta_{AB} = -\,\eta_{BA}$ is the 2-dimensional Levi-Civita tensor (with $\eta_{23} = 1$). Obviously this is a non-flat space. The corresponding Riemannian connection (\ref{rDS3}) yields a nontrivial curvature $\widehat{R}_B{}^A = \lambda\,\vartheta_B \wedge\vartheta^A$ of a 2-dimensional de Sitter space. The volume 4-form of the spacetime manifold reads $\eta = \vartheta^{\widehat 0}\wedge\vartheta^{\widehat 1}\wedge\vartheta^{\widehat 2}\wedge\vartheta^{\widehat 3} = {\frac {q^2}{2p^2}}k\wedge d\rho\wedge\underline{\eta}$. For the wave 1-form we find the remarkable relation
\begin{equation}
{}^*k = -\,k\wedge\underline{\eta}.\label{dualkDS}
\end{equation}
We will denote the geometrical objects on the transversal 2-space by underlining them; for example, a 1-form $\underline{\phi} = \phi_A\vartheta^A$. The Hodge duality on this space is defined as usual via ${}^{\underline{*}}\vartheta_A = \underline{\eta}_A = e_A\rfloor \underline{\eta} = \eta_{AB}\vartheta^B$. With the help of (\ref{dualkDS}), we can verify
\begin{equation}
{}^*(k\wedge\underline{\phi}) = k\wedge{}^{\underline{*}}\underline{\phi}.\label{kphiDS}
\end{equation}

The new 1-forms (\ref{THDS}) and (\ref{OMDS}) have the obvious transversality properties:
\begin{equation}
k\wedge{}^*\Theta = 0,\qquad k\wedge{}^*\Omega^\alpha = 0,\qquad 
k_\alpha\Omega^\alpha = 0.\label{kTOMDS}
\end{equation}
In accordance with (\ref{Wa0}) and (\ref{OMDS}), we have explicitly:
\begin{eqnarray}
\Theta &=& \vartheta^A\left({\frac 12}\widehat{D}_AU - \delta_{AB}W^B\right),\label{THDS2}\\
\Omega^A &=& \vartheta^B\left(\widehat{D}_BW^A + {\frac \lambda 2}\,U\delta_B^A\right).\label{OMDS2}
\end{eqnarray}
Here we denoted $\widehat{D}_A = e_A\rfloor\widehat{D}$. Applying the transversal differential to (\ref{THDS}), and making use of (\ref{OMDS}), we find 
\begin{equation}
\underline{d}\Theta = \Omega_\alpha\wedge\vartheta^\alpha.\label{dTHDS}
\end{equation}
In essence, this is equivalent to the Bianchi identity $DT^\alpha = R_\beta{}^\alpha\wedge\vartheta^\beta$ which is immediately checked by applying the covariant differential $D$ to (\ref{torDS}) and using (\ref{curDSPG}). Note that it is crucial to use (\ref{dkDS1}). 

A further refinement of the generalized wave ansatz will be considered in Sec.~\ref{Tansatz}.

\subsection{Irreducible decomposition of gravitational field strengths}

Irreducible parts of the torsion and the curvature are as follows. The second (trace) and third (axial trace) irreducible part of the torsion are trivial, ${}^{(2)}\!T^\alpha = 0$ and ${}^{(3)}\!T^\alpha = 0$, and the first (pure tensor) piece is nontrivial:
\begin{equation}
{}^{(1)}\!T^\alpha = T^\alpha = -\,k\wedge{\frac qp}k^\alpha\,\Theta.\label{torDS1}
\end{equation}
At the same time, the curvature pieces ${}^{(3)}\!R^{\alpha\beta} = {}^{(5)}\!R^{\alpha\beta} = 0$,
whereas 
\begin{equation}
{}^{(6)}\!R^{\alpha\beta} = \lambda\,\vartheta^\alpha\wedge\vartheta^\beta,\label{curDS6} 
\end{equation}
and for $I = 1,2,4$:
\begin{equation}
{}^{(I)}\!R^{\alpha\beta} = 2\,k\wedge {}^{(I)}\!\Omega^{[\alpha}k^{\beta]}{\frac qp}.\label{curDS124}
\end{equation}
Here ${}^{(1)}\!\Omega^\alpha + {}^{(2)}\!\Omega^\alpha + {}^{(4)}\!\Omega^\alpha = \Omega^\alpha$, and explicitly we have
\begin{eqnarray}
{}^{(1)}\!\Omega^\alpha &=& {\frac 12}\left(\Omega^\alpha - \vartheta^\alpha e_\beta\rfloor\Omega^\beta
+ \vartheta^\beta e^\alpha\rfloor\Omega_\beta\right),\label{OMDSi1}\\
{}^{(2)}\!\Omega^\alpha &=& {\frac 12}\left(\Omega^\alpha - \vartheta^\beta e^\alpha\rfloor
\Omega_\beta\right),\label{OMDSi2}\\
{}^{(4)}\!\Omega^\alpha &=& {\frac 12}\,\vartheta^\alpha e_\beta\rfloor\Omega^\beta.\label{OMDSi4}
\end{eqnarray}
The transversal components of these objects are constructed in terms of the irreducible pieces of the $2\times 2$ matrix $\widehat{D}_BW^A$: symmetric traceless part, skew-symmetric part and the trace, respectively. Using (\ref{OMDS2}), we derive ${}^{(I)}\!\Omega^A = {}^{(I)}\!\Omega^A{}_B\,\vartheta^B$, with
\begin{eqnarray}
{}^{(1)}\!\Omega^A{}_B &=& {\frac 12}\left(\widehat{D}_BW^A + \widehat{D}^AW_B - \delta^A_B
\,\widehat{D}_CW^C\right),\label{OMDSAB1}\\
{}^{(2)}\!\Omega^A{}_B &=& {\frac 12}\left(\widehat{D}_BW^A - \widehat{D}^AW_B\right),\label{OMDSAB2}\\
{}^{(4)}\!\Omega^A{}_B &=& {\frac 12}\,\delta^A_B\left(\widehat{D}_CW^C + \lambda\,U\right).\label{OMDSAB4}
\end{eqnarray}

One can demonstrate the following properties of these 1-forms:
\begin{eqnarray}\label{vOM}
\vartheta_\alpha\wedge{}^{(1)}\!\Omega^\alpha = 0,\quad \vartheta_\alpha\wedge{}^{(2)}\!\Omega^\alpha 
= \vartheta_\alpha\wedge\Omega^\alpha,\\
\vartheta_\alpha\wedge{}^{(4)}\!\Omega^\alpha = 0,\quad
e_\alpha\rfloor{}^{(1)}\!\Omega^\alpha = -\,e_\alpha\rfloor\Omega^\alpha,\\ 
e_\alpha\rfloor{}^{(2)}\!\Omega^\alpha = 0,\quad e_\alpha\rfloor{}^{(4)}\!\Omega^\alpha 
= 2e_\alpha\rfloor\Omega^\alpha,\label{eOM}\\
k_\alpha{}^{(1)}\!\Omega^\alpha = -\,{\frac 12}\,k\,e_\alpha\rfloor\Omega^\alpha,\quad
k_\alpha{}^{(2)}\!\Omega^\alpha = 0,\label{kOM}\\
k_\alpha{}^{(4)}\!\Omega^\alpha = {\frac 12}\,k\,e_\alpha\rfloor\Omega^\alpha,
\quad k\wedge{}^*{}^{(2)}\!\Omega^\alpha = 0,\\
k\wedge{}^*{}^{(1)}\!\Omega^\alpha = -\, k\wedge{}^*{}^{(4)}\!\Omega^\alpha = 
-\,{\frac 12}\,k^\alpha\,\vartheta_\beta\wedge{}^*\Omega^\beta.\label{kdOM}
\end{eqnarray}

\section{Field equations}\label{FE}

Let us now turn to the quadratic Poincar\'e gauge model with the general Lagrangian (\ref{LRT}), and allow for a nontrivial cosmological constant $\lambda_0$. 

Substituting the torsion (\ref{torDS1}) and the curvature (\ref{curDS6}), (\ref{curDS124}), into (\ref{HlaQT}) and (\ref{hR}), we find 
\begin{eqnarray}
h^\alpha &=& -\,k^\alpha Z{\frac qp},\\ 
h^{\alpha\beta} &=& \lambda b_6\eta^{\alpha\beta} + \lambda
\overline{b}_6\vartheta^\alpha\wedge\vartheta^\beta -\,2k^{[\alpha}Z^{\beta]}{\frac qp},\label{hDSZ}
\end{eqnarray}
where we introduced the 2-forms 
\begin{eqnarray}
Z &=& a_1\,{}^*\!\left(k\wedge\Theta\right) + \overline{a}_1\,k\wedge\Theta,\label{ZTDS}\\
Z^\alpha &=& \sum\limits_{I=1,2,4} \left[b_I\,{}^*\!\left(k\wedge{}^{(I)}\!\Omega^\alpha\right) 
+ \overline{b}_I\,k\wedge{}^{(I)}\!\Omega^\alpha\right].\label{ZRDS}
\end{eqnarray}
Making use of (\ref{kTOMDS}) and (\ref{vOM})-(\ref{kdOM}) we can show that
\begin{eqnarray}
k\wedge h^\alpha = 0,\qquad k\wedge {}^*h^\alpha = 0,\qquad k_\alpha h^\alpha = 0.\label{kha}
\end{eqnarray}
As a result, substituting (\ref{hDSZ}) into (\ref{qa}) and (\ref{qaR}), we find 
$q^{(T)}_\alpha = 0$ and 
\begin{eqnarray}
q^{(R)}_\alpha &=& 2\lambda\,{\frac qp}k_\alpha\bigl\{ -\,(b_4 + b_6)
\,{}^*\!k\,e_\beta\rfloor\Omega^\beta \nonumber\\
&& +\,(\overline{b}_2 - \overline{b}_6)\,k\wedge\vartheta_\beta\wedge
\Omega^\beta\bigr\}.\label{qqDS}
\end{eqnarray}
With an account of the properties (\ref{kha}), one can check that
\begin{eqnarray}
Dh_\alpha &=& -\,\widehat{D}\Bigl(k_\alpha Z{\frac qp}\Bigr),\label{dhaDS}\\
Dh_{\alpha\beta} &=& -\,\widehat{D}\Bigl(2k_{[\alpha}Z_{\beta]}{\frac qp}\Bigr)
+ \lambda b_6\eta_{\alpha\beta\mu}\wedge T^\mu \nonumber\\
&& +\,\lambda \overline{b}_6\left(T_\alpha\wedge\vartheta_\beta 
- T_\beta\wedge\vartheta_\alpha\right).\label{dhabDS}
\end{eqnarray}

The transversal nature of $\Theta$ and $\Omega^A$ leads to a further simplification. 
In particular, using (\ref{kphiDS}), we recast (\ref{ZTDS}) and (\ref{ZRDS}) into
\begin{equation}
Z = k\wedge\Xi,\qquad Z^A = k\wedge\Xi^A,\label{ZZADS}
\end{equation}
where we have introduced the 1-forms
\begin{eqnarray}
\Xi &=& a_1\,{}^{\underline{*}}\Theta + \overline{a}_1\,\Theta,\label{xiDS}\\
\Xi^A &=& \sum\limits_{I=1,2,4} \left[b_I\,{}^{\underline{*}}{}^{(I)}\!\Omega^A 
+ \overline{b}_I\,{}^{(I)}\!\Omega^A\right].\label{xiADS}
\end{eqnarray}

\begin{widetext}

\subsection{Wave equations}

After all these preparations, we are in a position to write down the gravitational field equations for the quadratic Poincar\'e gauge model (\ref{LRT}). Substituting the gravitational wave ansatz (\ref{cofDSPG})-(\ref{conDS}) into (\ref{ERT1}), we derive the first equation
\begin{eqnarray}
\left(3a_0\lambda - \lambda_0\right)\eta_\alpha + {\frac qp}k_\alpha\,{}^*\!k\,(e_\beta\rfloor
\Omega^\beta)\left[a_0 - 2\lambda\ell_\rho^2(b_4 + b_6)\right]\nonumber\\ \label{EPGDS1}
+ \,{\frac qp}k_\alpha\,k\wedge\Bigl\{\vartheta_\beta\wedge\Omega^\beta\Bigl[\overline{a}_0 +
2\lambda\ell_\rho^2(\overline{b}_2 - \overline{b}_6)\Bigr] - \underline{d}\,\Xi\Bigr\} = 0.
\end{eqnarray}
Contracting this with $k^\alpha$, we find the value of the constant parameter in the wave
ansatz:
\begin{equation}
\lambda = {\frac {\lambda_0}{3a_0}},\label{laDS}
\end{equation}
and with an account of (\ref{dualkDS}) and (\ref{xiDS}) we recast (\ref{EPGDS1}) into
\begin{eqnarray}
\left[a_0 - 2\lambda\ell_\rho^2(b_4 + b_6)\right]\vartheta_A\wedge{}^{\underline{*}}\Omega^A 
+ a_1\,\underline{d}\,{}^{\underline{*}}\!\Theta - \Bigl[\overline{a}_0 
+ \overline{a}_1 + 2\lambda\ell_\rho^2(\overline{b}_2 - \overline{b}_6)
\Bigr]\vartheta_A\wedge\Omega^A = 0.\label{EPDSW1}
\end{eqnarray}
The first two terms describe the parity-even model, whereas the last term accounts for 
the parity-odd sector.

Similarly, by gravitational wave ansatz (\ref{cofDSPG})-(\ref{conDS}) in (\ref{ERT2}), 
we obtain the second equation
\begin{eqnarray}
k_a\,{\frac qp}\,k\wedge\Big\{\left(a_0 + a_1 - 2\lambda\ell_\rho^2\,b_6\right)
\vartheta_B\wedge{}^{\underline{*}}\Theta + \Bigl(\overline{a}_0 + \overline{a}_1 
- 2\lambda\ell_\rho^2\,\overline{b}_6\Bigr)\vartheta_B\wedge\Theta 
- 2\ell_\rho^2\,\widehat{D}\,\Xi_B\Big\} = 0.\label{EPGDS2}
\end{eqnarray}
Note here that the $[ab]$ and $[AB]$ components in (\ref{ERT2}) are satisfied identically, and only the mixed $[aB]$ components give rise to the result (\ref{EPGDS2}). 

Equation (\ref{EPDSW1}) and the expression inside the curly bracket in (\ref{EPGDS2}) are both 2-forms on the 2-dimensional transversal space spanned by $x^A = (x^2, x^3)$, and thus (\ref{EPDSW1}) and (\ref{EPGDS2}) describe a system of three partial differential equations for the three variables $U = U(\sigma, x^B)$ and $W^A = W^A(\sigma, x^B)$. Substituting (\ref{THDS2}) and (\ref{OMDS2}), we recast (\ref{EPDSW1}) and (\ref{EPGDS2}) into the final tensorial form
\begin{eqnarray}
A_0(\widehat{D}_AW^A + \lambda U) + a_1(\widehat{D}_AW^A - {\frac 12}\,\widehat{\Delta}\,U)
- \overline{A}_0\,\eta^{AB}\widehat{D}_A\underline{W}{}_B &=& 0,\label{DSW1}\\
-\,A_1\,(\underline{W}{}_A - {\frac 12}\widehat{D}_AU) + \overline{A}_1\,\eta_{AB}
(W^B - {\frac 12}\underline{\widehat{D}}{}^BU) &&\nonumber\\
+ \,\ell_\rho^2(\overline{b}_1 - \overline{b}_2)\left[\widehat{D}_A(\eta^{BC}\widehat{D}_B
\underline{W}{}_C) + \eta_{AB}\underline{\widehat{D}}{}^B(\widehat{D}_CW^C + \lambda U)
\right] &&\nonumber\\
+ \,\ell_\rho^2(b_1 + b_4)\Big[-\,\widehat{\Delta}(\underline{W}{}_A - {\frac 12}
\widehat{D}_AU) + \lambda\,(\underline{W}{}_A - {\frac 12}\widehat{D}_AU) &&\nonumber\\
- \,\widehat{D}_A(\widehat{D}_BW^B + \lambda U) + \widehat{D}_A(\widehat{D}_BW^B 
- {\frac 12}\widehat{\Delta}\,U)\Big] &=& 0.\label{DSW2}
\end{eqnarray}
The 2-dimensional transversal space has the (anti)-de Sitter geometry and the corresponding covariant Laplacian reads
\begin{equation}\label{LapDS}
\widehat{\Delta} = \delta^{AB}\widehat{D}_A\widehat{D}_B = p^2\,\underline{\Delta},
\end{equation}
where $\underline{\Delta} = \delta^{AB}\partial_A\partial_B$ is the usual Laplace operator.
\end{widetext}
Note that $\overline{b}_4 = \overline{b}_2$. Here we denoted $\underline{W}{}_A = \delta_{AB}W^B$ and $\underline{\widehat{D}}{}^A = \delta^{AB}\widehat{D}_B$, and introduced the convenient abbreviations for the combinations of the coupling constants,
\begin{eqnarray}
A_0 &=& a_0 - 2\lambda\ell_\rho^2(b_4 + b_6),\label{A0}\\
\overline{A}{}_0 &=& \overline{a}_0 + \overline{a}_1 + 2\lambda\ell_\rho^2
(\overline{b}_2 - \overline{b}_6),\label{A2a}\\
A_1 &=& a_0 + a_1 + 2\lambda\ell_\rho^2(b_1 - b_6),\label{A1}\\
\overline{A}{}_1 &=& \overline{a}_0 + \overline{a}_1 + 2\lambda\ell_\rho^2
(\overline{b}_1 - \overline{b}_6).\label{A2}
\end{eqnarray}
The transversal covariant derivatives do not commute,
\begin{equation}
(\widehat{D}_A\widehat{D}_B - \widehat{D}_B\widehat{D}_A)W^C = \widehat{R}_{ABD}{}^C W^D 
= 2\lambda\,\delta^C_{[A}\underline{W}_{B]},\label{comDD}
\end{equation}
and we used this fact when deriving (\ref{DSW1}) and (\ref{DSW2}). Direct consequences
of (\ref{comDD}) are: 
\begin{eqnarray}
\eta^{BC}\widehat{D}_B\widehat{D}_C\underline{W}_A &=& \lambda\eta_{AB}W^B,\label{etaDD}\\
(\widehat{\Delta}\,\widehat{D}_A - \widehat{D}_A\,\widehat{\Delta})U &=& \lambda
\widehat{D}_AU.\label{comDU}
\end{eqnarray}

It is worthwhile to notice that the derivatives of $W^A$ appear in (\ref{DSW1})-(\ref{DSW2}) 
only in combinations
\begin{eqnarray}
\Omega &:=&  e_\alpha\rfloor\Omega^\alpha = \widehat{D}_AW^A + \lambda U,\label{OWU}\\
\Phi &:=&  {}^{\underline{*}}\!\underline{d}\,{}^{\underline{*}}\!\Theta 
= \widehat{D}_AW^A - {\frac 12}\,\widehat{\Delta}\,U,\label{PDU}\\
\overline{\Phi} &:=& {}^{\underline{*}}\!\underline{d}\,\Theta =
-\,\eta^{AB}\widehat{D}_A\underline{W}{}_B,\label{PDW}
\end{eqnarray}
which have a clear geometrical meaning in terms of the curvature and the torsion. 

The system (\ref{DSW1})-(\ref{DSW2}) always admits a nontrivial solution for arbitrary quadratic Poincar\'e gauge model with any choices of the coupling constants. There are some interesting special cases.

\subsection{Torsionless gravitational waves}

The torsion (\ref{torDS}) vanishes when $\Theta = 0$
which is realized, see (\ref{THDS}) and (\ref{THDS2}), for 
\begin{equation}
W^A = {\frac 12}\delta^{AB}\widehat{D}_BU.\label{notorDSW}
\end{equation}
Substituting this into (\ref{DSW1}), we find 
\begin{equation}
A_0\left\{\widehat{\Delta}\,U + 2\lambda\,U\right\} = 0,\label{notDS1}
\end{equation}
whereas (\ref{DSW2}) reduces to 
\begin{eqnarray}
\ell_\rho^2\,(\overline{b}_1 - \overline{b}_2)\,\eta_{AB}\,\underline{\widehat{D}}{}^B
\left\{\widehat{\Delta}\,U + 2\lambda\,U\right\}&& \nonumber\\
-\,\ell_\rho^2\,(b_1 + b_4)\,\widehat{D}_A\left\{\widehat{\Delta}\,U + 2\lambda\,U\right\} 
&=& 0.\label{notDS2}
\end{eqnarray}
Accordingly, we conclude that the well-known torsionless wave solution of GR with the
function $U$ satisfying 
\begin{equation}
p^2\,\underline{\Delta}\,U + 2\lambda\,U = 0\label{DSGR}
\end{equation}
is an exact solution of the generic quadratic Poincar\'e gauge gravity model. This is consistent with our earlier results on the torsion-free solutions in Poincar\'e gauge theory \cite{selected}. 

Moreover, the torsionless wave (\ref{notorDSW})-(\ref{notDS1}) represents a general solution for the purely torsion quadratic class of  Poincar\'e models, since this is the only configuration admitted by the system (\ref{DSW1})-(\ref{DSW2}) for $b_I = \overline{b}_I = 0$.

\subsection{Torsion gravitational waves}\label{Tansatz}

The torsion-free ansatz (\ref{notorDS}) can be generalized to 
\begin{equation}
W^A = {\frac 12}\,\delta^{AB}\widehat{D}_B(U + V) + {\frac {1}{2}}\,\eta^{AB}
\widehat{D}_B\overline{V},\label{pDSW}
\end{equation}
with $V \neq 0$. The two scalar functions $V = V(\sigma, x^A)$ and $\overline{V} = \overline{V}(\sigma, x^A)$ define the non-Riemannian piece of the connection, stemming from torsion:
\begin{eqnarray}
\Theta &=& -\,{\frac 12}\left(\underline{d}\,V 
+ {}^{\underline{*}}\!\underline{d}\,\overline{V}\right)\nonumber\\
&=& -\,{\frac 12}\,\vartheta^A\left(\widehat D_AV - 
\eta_{AB}\underline{\widehat{D}}{}^B\overline{V}\right)\,.
\end{eqnarray}
For the above choice, the metric and torsion contributions to the connection are described in a rather symmetric way, in terms of the three potentials ($U,V,\overline{V})$. In particular, we find for (\ref{OWU})-(\ref{PDW}):
\begin{eqnarray}\label{OV}
\Omega = {\frac 12}\left(\widehat{\Delta}\,V + \widehat{\Delta}\,U + 2\lambda U\right),\\
\Phi = {\frac 12}\widehat{\Delta}\,V,\qquad \overline{\Phi} = {\frac 12}\widehat{\Delta}
\,\overline{V}.\label{PV}
\end{eqnarray}

Substituting (\ref{pDSW}) into (\ref{DSW1}) and (\ref{DSW2}), we derive
\begin{widetext}
\begin{eqnarray}
A_0\Omega + a_1\Phi + \overline{A}{}_0\overline{\Phi} = 0,\label{pDSW1}
\end{eqnarray}
\begin{eqnarray}
\widehat{D}_A\Big\{ -\,{\frac 12}A_1V - {\frac 12}\overline{A}_1\overline{V} - \ell_\rho^2
(b_1 + b_4)\,\Omega -  \ell_\rho^2(\overline{b}_1 - \overline{b}_2)\overline{\Phi}\Big\} 
&& \nonumber\\
+\,\eta_{AB}\underline{\widehat{D}}{}^B\Big\{ -\,{\frac 12}A_1V + {\frac 12}\overline{A}_1
\overline{V} - \ell_\rho^2(b_1 + b_4)\,\overline{\Phi} 
+ \ell_\rho^2(\overline{b}_1 - \overline{b}_2)\Omega\Big\} &=& 0.\label{pDSW2}
\end{eqnarray}
One needs to pay attention to the noncommutativity of the covariant derivatives and use 
(\ref{comDD})-(\ref{comDU}). 

As a result, we obtain the system of the three linear second order differential equations
for the three functions $U, V, \overline{V}$:
\begin{eqnarray}
A_0(\widehat{\Delta}\,V + \widehat{\Delta}\,U + 2\lambda U) + a_1\widehat{\Delta}
\,V + \overline{A}{}_0\widehat{\Delta}\,\overline{V} &=& 0,\label{DSV1}\\
-\,\ell_\rho^2(b_1 + b_4)(\widehat{\Delta}\,V + \widehat{\Delta}\,U + 2\lambda U) - A_1V 
-\,\ell_\rho^2(\overline{b}_1 - \overline{b}_2)\widehat{\Delta}\,\overline{V} 
- \overline{A}{}_1\overline{V} &=& 0,\label{DSV2}\\
\ell_\rho^2(\overline{b}_1 - \overline{b}_2)(\widehat{\Delta}\,V + \widehat{\Delta}
\,U + 2\lambda U) + \overline{A}{}_1V 
-\,\ell_\rho^2(b_1 + b_2)\widehat{\Delta}\,\overline{V} - 
A_1\overline{V} &=& 0.\label{DSV3}
\end{eqnarray}
\end{widetext}

\subsection{Solution for potentials}

Before starting the analysis of solutions, one can notice that the system (\ref{DSV2}) and (\ref{DSV3}) is actually not equivalent to the original equation (\ref{pDSW2}). Indeed, by taking the covariant divergence (applying $\widehat{D}^A$) and by taking the covariant curl (applying $\eta^{AB}\widehat{D}_B$) of (\ref{pDSW2}), we derive the pair of equations where on the right-hand sides of (\ref{DSV2}) and (\ref{DSV3}) one finds not zeros but arbitrary functions, say, $\alpha(\sigma, x^A)$ and $\beta(\sigma, x^A)$, which are {\it harmonic}, in the sense that they both satisfy equations $\widehat{\Delta}\alpha = \widehat{\Delta}\beta = 0$. However, one then immediately notices that with the help of redefinitions
\begin{eqnarray}
V &\longrightarrow& V + v,\qquad \widehat{\Delta}v = 0,\label{gaugeV}\\
\overline{V} &\longrightarrow& \overline{V} + \overline{v},\qquad \widehat{\Delta}\overline{v} = 0,\label{gaugeVT}
\end{eqnarray}
we can always remove these nontrivial right-hand sides and come to the system (\ref{DSV2}) and (\ref{DSV3}). 

In other words, a solution of the system (\ref{DSV1})-(\ref{DSV3}) admits the {\it gauge} transformation (\ref{gaugeV})-(\ref{gaugeVT}), under which the potentials $V$ and $\overline{V}$ can be shifted by arbitrary harmonic functions. Such gauge transformed potentials are of course still solutions of the Poincar\'e gauge field equations (\ref{pDSW1}) and (\ref{pDSW2}). What is important, however, the curvature and the torsion remain invariant under the redefinition (\ref{gaugeV})-(\ref{gaugeVT}) of potentials: (\ref{OV}) and (\ref{PV}) obviously are not affected by the arbitrary harmonic functions. 

Now, as a first step, we substitute $(\widehat{\Delta}\,V + \widehat{\Delta}\,U + 2\lambda U)$ from (\ref{DSV1}) into (\ref{DSV2}) and (\ref{DSV3}). The resulting system reads
\begin{eqnarray}
&&\ell_\rho^2\widehat{\Delta}\left\{a_1(b_1 + b_4)V + [- A_0(\overline{b}_1 - 
\overline{b}_2) + \overline{A}_0(b_1 + b_4)]\overline{V}\right\}\nonumber\\
&&-\,A_0A_1V - A_0\overline{A}_1\overline{V} = 0, \label{DSVa}\\
&& \ell_\rho^2\widehat{\Delta}\left\{a_1(\overline{b}_1 - \overline{b}_2)V +
[A_0(b_1 + b_2) + \overline{A}_0(\overline{b}_1 - \overline{b}_2)]\overline{V}
\right\} \nonumber\\
&&-\,A_0\overline{A}_1V + A_0A_1\overline{V} = 0.\label{DSVb}
\end{eqnarray}
After solving this system, we can use the potentials $V$ and $\overline{V}$ to substitute them into (\ref{DSV1}) which then becomes an inhomogeneous differential equation for the metric potential $U$:
\begin{equation}
A_0(\widehat{\Delta}\,U + 2\lambda U) = -\,(a_1 + A_0)\widehat{\Delta}\,V 
- \overline{A}{}_0\widehat{\Delta}\,\overline{V}.\label{DUV}
\end{equation}
For the parity-even models with $\overline{a}_I = 0, \overline{b}_I = 0$, hence $\overline{A}_0 = 0$ and $\overline{A}_1 = 0$, the system (\ref{DSVa})-(\ref{DSVb}) reduces to the two uncoupled equations
\begin{eqnarray}
a_1(b_1 + b_4)\ell_\rho^2\widehat{\Delta}\,V - A_0A_1V &=& 0,\label{EDS1}\\
(b_1 + b_2)\ell_\rho^2\widehat{\Delta}\,\overline{V} + 
A_1\overline{V} &=& 0,\label{EDS2}
\end{eqnarray}
recovering the result of \cite{BC5}.

To analyse the system (\ref{DSVa})-(\ref{DSVb}), let us rewrite it in matrix form
\begin{equation}
\widehat{\Delta}\,{\cal V} - M\,{\cal V} = 0,\qquad 
M := {\frac {A_0}{\ell_\rho^2}}\,F,\label{DSVc}
\end{equation}
where we combined the potentials into a single object, a ``2-vector'' ${\cal V} = \left(\begin{array}{c} V \\ \overline{V}\end{array}\right)$, and the $2\times 2$ matrix $F = K^{-1}N$ is constructed from 
\begin{eqnarray}
K &=& \left(\begin{array}{c|c} a_1(b_1 + b_4) & \overline{A}_0(b_1 + b_4) 
- A_0(\overline{b}_1 - \overline{b}_2)\\ \hline a_1(\overline{b}_1 
- \overline{b}_2) & A_0(b_1 + b_2) + \overline{A}_0(\overline{b}_1 
- \overline{b}_2)\end{array}\right),\nonumber\\
N &=& \left(\begin{array}{c|c} A_1 & \overline{A}_1\\ \hline
\overline{A}_1 & - A_1 \end{array}\right).\label{KN}
\end{eqnarray}
One immediately notices the simple structure of the matrix $N$ which is manifest
in the properties
\begin{equation}\label{N2}
N^2 = (A_1^2 + \overline{A}{}_1^2)\left(\begin{array}{c|c} 1 & 0 \\ 
\hline 0 & 1 \end{array}\right),\quad \det\,N = - (A_1^2 + \overline{A}{}_1^2).
\end{equation}

One can solve the matrix differential equation (\ref{DSVc}) by diagonalizing this system. To achieve this, one needs to find the eigenvalues of the matrix $M$ and to construct the corresponding eigenvectors. Let $m^2$ be an eigenvalue of the matrix $M$. It is determined from the corresponding characteristic equation  $\det (M-m^2)=0$ which has the meaning of the dispersion relation for the mass:
\begin{equation}\label{meq}
\ell_\rho^4m^4\det K + \ell_\rho^2m^2A_0{\rm tr}(NK) - A_0^2(A_1^2 + \overline{A}{}_1^2) = 0.
\end{equation}
The coefficients of the quadratic equation (\ref{meq}) are constructed from the 
coupling constants of the gauge gravity model. From (\ref{KN}) we have explicitly:
\begin{eqnarray}\label{detK}
\det K &=& a_1A_0\left[(b_1 + b_4)(b_1 + b_2) + (\overline{b}_1 - \overline{b}_2)^2\right],\\
{\rm tr}(NK) &=& (a_1A_1 +\overline{A}{}_0\overline{A}{}_1)(b_1 + b_4) - A_0A_1(b_1 + b_2)\nonumber\\
&& +\,(a_1\overline{A}_1 - A_0\overline{A}_1 - \overline{A}_0A_1)(\overline{b}_1 
- \overline{b}_2).\label{trM}
\end{eqnarray}

For the parity-even models with $\overline{a}_I = 0, \overline{b}_I = 0$, hence $\overline{A}_0 = 0$ and $\overline{A}_1 = 0$, the dispersion equation (\ref{meq}) reduces to 
\begin{eqnarray}
\left[\ell_\rho^2m^2a_1(b_1 + b_4) - A_0A_1\right]\times\nonumber\\
\times \left[\ell_\rho^2m^2A_0(b_1 + b_2) + A_0A_1\right] = 0,\label{meqE}
\end{eqnarray}
and hence we recover the result (\ref{EDS1})-(\ref{EDS2}).

General case with parity-odd terms in the Lagrangian is more complicated. No obvious
simplification of (\ref{meq}) is visible. 

Having found the eigenvalues $m_1^2$ and $m_2^2$ of the mass matrix $M$ as the two roots of the quadratic equation (\ref{DSVc}), one can construct the matrix $P$ that transforms $M$ to its diagonal form. For $M_{12}\ne 0$, the latter reads
\begin{equation}\label{P}
P = \left(\begin{array}{cc} -M_{12} & -M_{12}\\ M_{11}-m_1^2 & M_{11}-m_2^2\end{array}\right).
\end{equation}
Multiplying Eq. (\ref{DSVc}) by $P^{-1}$, one then obtains
\begin{equation}
\widehat\Delta{\cal V}' - M'{\cal V}' = 0\,,
\end{equation}
where 
\begin{equation}\label{M1}
M' := P^{-1}MP = \left(\begin{array}{cc} m_1^2  &  0 \\ 0 & m_2^2\end{array}\right)\,,
\end{equation}
and ${\cal V}'$ is the eigenvector of $M$, corresponding to the eigenvalues $m_1^2$ and $m_2^2$:
\begin{eqnarray}
{\cal V}' &=& \left(\begin{array}{c} {\cal V}'_1 \\ {\cal V}'_2\end{array}\right) 
= P^{-1}{\cal V}\nonumber\\
&=& {\frac 1{\det P}}\left(\begin{array}{c} (M_{11} - m_2^2)V + M_{12}\overline{V} \\
- (M_{11} - m_1^2)V - M_{12}\overline{V}\end{array}\right).
\end{eqnarray}

Recalling $\widehat\Delta = p^2\underline{\Delta}$, we thus recast the system of the field equations (\ref{DSVa}) and (\ref{DSVb}) into a diagonal form
\begin{equation}
p^2\underline{\Delta}{\cal V}'_n - m_n^2{\cal V}'_n=0\,,
\end{equation}
with $n = 1, 2$. The solutions for ${\cal V}'_n$ are given in terms of the hypergeometric functions ${}_2F_1(a,b,c,z)$, see \cite{BC5}. Similar construction exists in the case $M_{21}\neq 0$.

Now, we can return to (\ref{DUV}) to find the solution for $U$. Each solution for ${\cal V}_n'$ defines the corresponding solution 
\begin{equation}
{\cal V} = P {\cal V}'\,
\end{equation}
of (\ref{DSVc}). Inserting these solutions for $V$ and $\overline{V}$ on the right hand side of (\ref{DUV}), this equation becomes an inhomogeneous differential equation for $U$. Its general solution is given as a general solution of the homogeneous equation plus a particular solution of the inhomogeneous equation, $U=U_\text{h}+U_\text{p}$. Note that $U_\text{h}$ coincides with the general vacuum solution of GR, see (\ref{DSGR}). The solution for $U$ obtained by choosing $U_\text{h}=0$ has a very interesting interpretation. Indeed, in that case $U$ reduces just to the particular solution $U_\text{p}$, the form of which is completely determined by the torsion potentials $(V,\overline{V})$. A similar mechanism was found also in the parity even sector \cite{BC5}. Clearly, there are many other solutions for $U_\text{h}$, and consequently, for $U$. In each of them, the influence of torsion on the metric is quite clear.

\subsection{Masses of the torsion modes}

In order to get a deeper understanding of the role of the torsion in our gravitational wave solution, it is important to examine the mass spectrum of the associated torsion modes. Having found the matrix $F=K^{-1}N$ with the help of (\ref{KN}), the solutions of the characteristic equation (\ref{meq}) can be conveniently represented in terms of the matrix $f=(\det K)F$ as
\begin{equation}
m^2_\pm = {\frac {A_0}{2\ell^2\det K}}\left({\rm tr} f\pm
\sqrt{({\rm tr} f)^2 - 4\det f}\right)\,.\label{4.66}
\end{equation}
This is an exact formula for the mass eigenvalues $m^2_\pm$ associated to the gravitational wave. It is worthwhile to notice that ${\rm tr} f = -\,{\rm tr}(NK)$, and $\det f = (\det N)(\det K)$.

The particle spectrum of PGT has been calculated only with respect to the Minkowski background \cite{Hay2,Nev1,Nev2,Sez1,Sez2,Kuh}, and never for the (anti)-de Sitter spacetime. Accordingly, we can compare the result (\ref{4.66}) with those existing in the literature only for the values of $m^2_{\pm}$ in the limit of the vanishing cosmological constant. In the limit of $\lambda\rightarrow 0$, we have
\begin{eqnarray}
{\rm tr} f &=& -\,\Big[a_1(a_0 + a_1) + (\overline{a}_0 + \overline{a}_1)^2\Big]
(b_1 + b_4)\nonumber\\
&& +\,a_0(a_0 + a_1)(b_1 + b_2) + 2a_0(\overline{a}_0 + \overline{a}_1)
(\overline{b}_1 - \overline{b}_2)\,,\nonumber\\
\det f&=&-\,a_0a_1\Big[(a_0 + a_1)^2+(\overline{a}_0 + \overline{a}_1)^2\Big]\times\nonumber\\
&& \times\Big[(b_1 + b_2)(b_1 + b_4) + (\overline{b}_1 - \overline{b}_2)^2\Big]\,,\nonumber\\
\det K &=& a_0a_1\Big[(b_1 + b_2)(b_1 + b_4) + (\overline{b}_1 - \overline{b}_2)^2\Big]\,.
\label{4.67}
\end{eqnarray}

As a first test, we apply the formula (\ref{4.66}) to the parity even sector of PGT. One can straightforwardly see that the corresponding values of $m^2_\pm$ coincide with the masses of the spin-$2^\pm$ torsion modes, known from the literature \cite{Hay2}; compare also with \cite{BC5}. This is consistent with (\ref{meqE}).

A more complete verification can be done by comparing (\ref{4.66}) with the recent work of Karananas \cite{Karananas}, which presently offers the only existing calculation of the complete mass spectrum for the most general PGT with both parity even and parity odd sectors included. A comparison of the Lagrangian (5) of ref. \cite{Karananas} with our expression (\ref{LRT}) is straightforward, although one should be careful since the paper \cite{Karananas} contains numerous misprints. As a result, we establish the following relations between our and Karananas' coupling constants (we use the notation $t_0$ instead of Karananas' $\lambda$ to distinguish it from our cosmological term):
\begin{eqnarray}\label{aa0}
a_0 &=& 2\kappa c \,t_0,\qquad \overline{a}_0 = 0,\\
a_1 &=& 2\kappa c \,(- t_1 - t_0),\label{aa1}\\ 
a_2 &=& 4\kappa c \,(- t_3 + t_0),\label{aa2}\\ 
a_3 &=& \kappa c \,(- t_2 + t_0),\label{aa3}\\
\overline{a}_1 &=& 4\kappa c \,t_5,\label{aa4}\\
\overline{a}_2 &=& \overline{a}_3 = 2\kappa c \,t_4.\label{aa5}
\end{eqnarray}
\begin{eqnarray}
b_1 &=& 4\rho\left(-r_1 + r_3\right),\label{bb1}\\
b_2 &=& 4\rho\left(-r_3\right),\label{bb2}\\
b_3 &=& 4\rho\left(-r_2 + r_3\right),\label{bb3}\\
b_4 &=& 4\rho\left(-r_1 + r_3 - r_4\right),\label{bb4}\\
b_5 &=& 4\rho\left(-r_3 - r_5\right),\label{bb5}\\
b_6 &=& 4\rho\left(-r_1 + r_3 - 3r_4\right),\label{bb6}
\end{eqnarray}
\begin{eqnarray}
\overline{b}_1 &=& \rho\left(-r_7 + 3r_8\right),\label{bbt1}\\
\overline{b}_2 &=& \overline{b}_4 = \rho\left(-r_7 - r_8\right),\label{bbt2}\\
\overline{b}_3 &=& \overline{b}_6 = \rho\left(4r_6 - r_7 - r_8\right),\label{bbt3}\\
\overline{b}_5 &=& \rho\left(3r_7 - r_8\right).\label{bbt4}
\end{eqnarray}
Substituting the expressions for $a_I, b_I$ and $\overline{a}_I, \overline{b}_I$ into (\ref{4.67}), one finds that the resulting values of $m^2_\pm$ in (\ref{4.66}) exactly reproduce the result (A.3.5) of Karananas' paper \cite{Karananas} (after correcting a number of his misprints), which displays the spin-$2^\pm$ torsion modes. 

Thus, we conclude that the massive spin-$2^\pm$ torsion modes turn out to be an essential ingredient of our gravitational wave, in the sense that these massive torsion modes determine the structure of the wave profile encoded in the functions $V, \overline{V}$ and $U$. This is a remarkable result if one recalls that the particle spectrum of PGT is derived from the linearized equations of motion, whereas our gravitational waves are exact solutions of the full nonlinear field equations.

\section{Discussion and conclusion}\label{DC}

In this paper, we have found a family of the exact vacuum solutions of the most general PGT model with all possible parity even and parity odd linear and quadratic invariants in the Lagrangian (\ref{LRT}), and with a nontrivial cosmological constant $\lambda_0\neq 0$. This family represents generalized plane waves with torsion, propagating on the (anti)-de Sitter background. The present paper extends the results obtained recently in \cite{BC5,PGW}.

The underlying construction can be understood as a generalization of the Kerr-Schild-Kundt ansatz from the Riemannian to the Riemann-Cartan geometry of PGT. An essentially new element in this extended formalism is the ansatz for the local Lorentz connection $\Gamma_\alpha{^\beta}$, the radiation piece of which is constructed in terms of the null covector field $k$. The generalized plane wave ansatz (\ref{cofDSPG})-(\ref{conDS}) ensures that the torsion 2-form $T^\alpha$ and the radiation piece of the curvature 2-form $S^{\alpha\beta} := R^{\alpha\beta} - \lambda\vartheta^\alpha\wedge\vartheta^\beta$ satisfy the radiation conditions
\begin{eqnarray}
k\wedge{}^\ast\!T^\alpha = 0,\qquad k\wedge{}^\ast\!S^{\alpha\beta} = 0,\label{kTW}\\
k\wedge T^\alpha = 0,\qquad k\wedge S^{\alpha\beta} = 0,\label{kRW}\\
T^\alpha\wedge{}^\ast\!T^\beta = 0,\qquad  
S^{\alpha\beta}\wedge{}^\ast\!S^{\rho\sigma} = 0.\label{kTRW}
\end{eqnarray}
These relations represent an extension of the well-known Lichnerowicz criterion for identifying gravitational waves \cite{Lich} (see also \cite{vdz}), based on analogy with the electromagnetic waves, to the framework of the PGT.

In the limit of vanishing torsion, the generalized plane waves with torsion reduce to the family of the Riemannian $pp$ waves on the (anti)-de Sitter background. The $pp$ waves are classified as solutions of Petrov type $N$, since the corresponding Weyl tensor satisfies the special algebraic condition $k^\alpha C_{\alpha\beta\mu\nu}=0$, see \cite{exact,griff2}. This criterion can be naturally extended to a Riemann-Cartan geometry of PGT as
\begin{equation}
k^\alpha\,{}^{(1)}\!R_{\alpha\beta\mu\nu} = 0\,,\label{5.71}
\end{equation}
where ${}^{(1)}\!R_{\alpha\beta\mu\nu}$ is the first irreducible part of the curvature tensor, see \cite{BC5,PGW}. The validity of (\ref{5.71}) for the generalized plane waves with torsion confirms that they are also of type $N$.

The spacetime torsion is an essential ingredient of the generalized gravitational wave solution; its dynamical characteristics are described by the two potentials $V$ and $\overline{V}$, satisfying the matrix equation (\ref{DSVc}). The mass matrix $M$ is of particular importance for the physical interpretation of the torsion. We demonstrate that, in the limit of $\lambda\rightarrow 0$, the eigenvalues of $M$ coincide with the values of the mass square  the spin-$2^\pm$ torsion modes, identified in the work of Karananas \cite{Karananas}. Generically, wave front profile of a generalized plane wave with torsion is thus determined by two spin-$2$ massive torsion modes and the massless graviton, produced by the third, coframe potential $U$ (which enters the spacetime metric).

It is interesting to note that there exist particular solutions for which the metric potential is completely determined by the torsion. For such solutions, the motion of a spinless test particle is effectively determined by the spacetime torsion.

The results obtained in this work were checked with the help of the computer algebra systems {\it Reduce} and {\it Mathematica}.

\begin{acknowledgments}

We dedicate this paper to Friedrich Hehl on the occasion of his 80th birthday, in appreciation of his contribution to the development of gauge theories of gravity.

We are grateful to Georgios Karananas for the clarifying comments on his study of the particle spectrum in the general quadratic Poincar\'e gravity models. YNO thanks Metin Gurses for the correspondence, kindly informing about his recent work. This work was partially supported by the Serbian Science Foundation (Grant No. 171031) and by the Russian Foundation for Basic Research (Grant No. 16-02-00844-A).

\end{acknowledgments}

\bigskip

\appendix

\section{Irreducible decomposition of the torsion and curvature}\label{irreducible}

The torsion 2-form can be decomposed into the three 
irreducible pieces, $T^{\alpha}={}^{(1)}T^{\alpha} + {}^{(2)}T^{\alpha} + 
{}^{(3)}T^{\alpha}$, where
\begin{eqnarray}
{}^{(2)}T^{\alpha}&=& {\frac 13}\vartheta^{\alpha}\wedge (e_\nu\rfloor 
T^\nu),\label{iT2}\\
{}^{(3)}T^{\alpha}&=& {\frac 13}e^\alpha\rfloor(T^{\nu}\wedge
\vartheta_{\nu}),\label{iT3}\\
{}^{(1)}T^{\alpha}&=& T^{\alpha}-{}^{(2)}T^{\alpha} - {}^{(3)}T^{\alpha}.
\label{iT1}
\end{eqnarray}
The Riemann-Cartan curvature 2-form is decomposed $R^{\alpha\beta} = 
\sum_{I=1}^6\,{}^{(I)}\!R^{\alpha\beta}$ into the 6 irreducible parts 
\begin{eqnarray}
{}^{(2)}\!R^{\alpha\beta} &=& -\,{}^*(\vartheta^{[\alpha}\wedge
\overline{\Psi}{}^{\beta]}),\label{curv2}\\
{}^{(3)}\!R^{\alpha\beta} &=& -\,{\frac 1{12}}\,{}^*(\overline{X}
\,\vartheta^\alpha\wedge\vartheta^\beta),\label{curv3}\\
{}^{(4)}\!R^{\alpha\beta} &=& -\,\vartheta^{[\alpha}\wedge\Psi^{\beta]},
\label{curv4}
\end{eqnarray}
\begin{eqnarray}
{}^{(5)}\!R^{\alpha\beta} &=& -\,{\frac 12}\vartheta^{[\alpha}\wedge e^{\beta]}
\rfloor(\vartheta^\gamma\wedge X_\gamma),\label{curv5}\\
{}^{(6)}\!R^{\alpha\beta} &=& -\,{\frac 1{12}}\,X\,\vartheta^\alpha\wedge
\vartheta^\beta,\label{curv6}\\
{}^{(1)}\!R^{\alpha\beta} &=& R^{\alpha\beta} -  
\sum\limits_{I=2}^6\,{}^{(I)}R^{\alpha\beta},\label{curv1}
\end{eqnarray}
where 
\begin{eqnarray}
X^\alpha := e_\beta\rfloor R^{\alpha\beta},\quad X := e_\alpha\rfloor X^\alpha,
\label{WX1}\\
\overline{X}{}^\alpha := {}^*(R^{\beta\alpha}\wedge\vartheta_\beta),\quad
\overline{X} := e_\alpha\rfloor \overline{X}{}^\alpha,\label{WX2}
\end{eqnarray}
and 
\begin{eqnarray}
\Psi_\alpha &:=& X_\alpha - {\frac 14}\,\vartheta_\alpha\,X - {\frac 12}
\,e_\alpha\rfloor (\vartheta^\beta\wedge X_\beta),\label{Psia}\\
\overline{\Psi}{}_\alpha &:=& \overline{X}{}_\alpha - {\frac 14}\,\vartheta_\alpha
\,\overline{X} - {\frac 12}\,e_\alpha\rfloor (\vartheta^\beta\wedge 
\overline{X}{}_\beta)\label{Phia}.
\end{eqnarray}

Directly from the definitions (\ref{iT2})-(\ref{iT1}) and (\ref{curv2})-(\ref{curv1}),
one can prove the relations
\begin{eqnarray}\label{T23}
&T^\alpha\wedge{}^{(2)}T_\alpha = T^\alpha\wedge{}^{(3)}T_\alpha = {}^{(2)}T^\alpha\wedge{}^{(3)}T_\alpha,&\\
&R^{\alpha\beta}\wedge{}^{(2)}\!R_{\alpha\beta} = R^{\alpha\beta}\wedge{}^{(4)}\!R_{\alpha\beta} 
= {}^{(2)}\!R^{\alpha\beta}\wedge{}^{(4)}\!R_{\alpha\beta},&\nonumber\\ &{}&\label{R24} \\ 
&R^{\alpha\beta}\wedge{}^{(3)}\!R_{\alpha\beta} = R^{\alpha\beta}\wedge{}^{(6)}\!R_{\alpha\beta} 
= {}^{(3)}\!R^{\alpha\beta}\wedge{}^{(6)}R_{\alpha\beta},&\nonumber\\ &{}&\label{R36}
\end{eqnarray}
whereas $T^\alpha\wedge{}^{(1)}T_\alpha = {}^{(1)}T^\alpha\wedge{}^{(1)}T_\alpha$ and 
$R^{\alpha\beta}\wedge{}^{(1)}\!R_{\alpha\beta} = {}^{(1)}\!R^{\alpha\beta}\wedge{}^{(1)}
\!R_{\alpha\beta}$ and $R^{\alpha\beta}\wedge{}^{(5)}\!R_{\alpha\beta} = {}^{(5)}\!R^{\alpha\beta}
\wedge{}^{(5)}\!R_{\alpha\beta}$.

\end{document}